\documentclass[pre,twocolumn]{revtex4}
\usepackage[english]{babel}
\usepackage{amsmath}
\usepackage{amssymb}
\usepackage{epsfig}

\begin{document}

\title{Vortex knots on three-dimensional lattices of nonlinear oscillators coupled by space-varying links}
\author{Victor P. Ruban}
\email{ruban@itp.ac.ru}
\affiliation{Landau Institute for Theoretical Physics, RAS,
Chernogolovka, Moscow region, 142432 Russia} 
\date{\today}

\begin{abstract}
Quantized vortices in a complex wave field described by a defocusing nonlinear Schr\"odinger 
equation with a space-varying dispersion coefficient are studied theoretically
and compared to vortices in the Gross-Pitaevskii model with external potential.
A discrete variant of the equation is used to demonstrate numerically that vortex knots 
in three-dimensional arrays of oscillators coupled by specially tuned weak links 
can exist for as long times as many tens of typical vortex turnover periods.
\end{abstract}

\maketitle
%%%%%%%%%%%%%%%%%%%%%%%%%%%%%%%%%%%%%%%%%%%%%%%%%%%%%%%%%%%%%%%%%%%%%%%%%%%

\section{Introduction}

Vortices of different nature are ubiquitous objects of high interest in hydrodynamics, optics, 
and condensed-matter physics \cite{Saffman,Pismen, Donnelly}. In particular,
quantized vortices are persistent ``soft'' excitations in nonlinear wave systems 
described by a complex order parameter. Accordingly, some simplified mathematical 
models were suggested to study the phenomenon theoretically. A famous example of that is 
given by the Gross-Pitaevskii equation (GPE) for a dilute Bose-Einstein condensate (BEC) 
of cold atoms \cite{PS1,PS2,KFC2015}. 
In general, quantized vortices are not necessary involving the Plank's constant. 
They exist in classical nonlinear wave fields as well, since quasi-monochromatic waves
in many cases are described by the defocusing nonlinear  Schr\"odinger equation (NLSE), 
though typically in moving frames of reference.
The static and dynamical properties of quantized vortices were extensively studied
theoretically, numerically, and experimentally 
(see reviews \cite{FS2001,F2009,PGK2004,Kom2007,ParkerBar,WAB}).

In three-dimensional (3D) space, geometry and topology of vortices can be quite rich. 
In particular, vortices can form closed rings and simple filaments. 
Such structures were investigated, e.~g., in 
Refs.~\cite{SF2000,R2001,AR2001,GP2001,AR2002,RBD2002,AD2003,AD2004,D2005,Kelvin_waves,
ring_instability,v-2015,BWTCFCK2015,R2017-2,R2017-3,reconn-2017,top-2017,WBTCFCK2017,TWK2018,
WBB2014,talley,R2018PoF,ISS2019}. 
More complicated configurations as knots and links have also attracted much 
attention and have been a subject for many theoretical and experimental works 
\cite{RSB999,MABR2010,POB2012,KI2013,POB2014,LMB016,KKI2016,R2018-1,R2018-2,R2018-3,
hall1,hall2,Maucher,VPK2016,KS2018}. 
Especially interesting are long-lived knotted or linked vortex structures preserving 
their topology over many typical vortex turnover times.
Recently within GPE model it has been numerically found that spatial confinement of BECs 
is able to enhance lifetimes of simplest torus knots and links more than by an order 
of magnitude as compared to the lifetimes on a uniform background \cite{R2018-4,TRK2019}. 
At a mathematical level, a nonuniform equilibrium density profile for a BEC arises 
due to the trap potential, while the dispersive term of GPE (which is inversely proportional 
to the atomic mass) remains homogeneous. 
But it is worth pointing out that there exists another way to introduce non-uniformity,
namely by allowing the dispersive terms in a NLSE to have space-varying coefficients. 
To the best of the author's knowledge, an influence of variable dispersion on quantized vortices 
in three dimensions was not investigated previously. Here this gap begins to be partly filled. 

We first consider a weakly nonlinear field of nonuniformly coupled classical oscillators 
described by their normal complex variables $a({\bf r},t)=A({\bf r},t)\exp(-i\omega_0 t)$.
In the mathematically simplest variant, an appropriate  equation of 
motion for the complex envelope $A({\bf r},t)$ is
\begin{equation}
i(A_t+\gamma A)=-\frac{1}{2}\nabla\cdot (F({\bf r})\nabla A) +g|A|^2A,
\label{A_eq}
\end{equation}
where $\gamma>0$ is a small linear damping rate,  $g$ is a nonlinear coefficient,
positive for definiteness 
(usually in physical systems we have a  relatively small nonlinear frequency shift,
so $g|A|^2\ll \omega_0$), and $F({\bf r})$ is a given scalar function of spatial coordinates 
(in this work we do not consider the more complicated 
though even more rich case when $F$ is a space-dependent matrix). 

To have in mind a definite example, let us consider the following dynamical system:
\begin{equation}
\dot q_n=\frac{\partial {\mathsf H}}{\partial p_n},\qquad 
-(\dot p_n+2\gamma p_n)=\frac{\partial {\mathsf H}}{\partial q_n},
\end{equation}
where $n$ is a discrete (multi)index, $q_n(t)$ is a canonical coordinate, and 
$p_n(t)$ is a canonical momentum. We assume the Hamiltonian function ${\mathsf H}$ to be of the form
corresponding to coupled oscillators each distorted by a quartic potential
(a kind of Klein-Gordon lattice),
\begin{equation}
{\mathsf H}=\sum_n\Big[\frac{\omega_0}{2}(q_n^2+p_n^2)+\frac{\varkappa}{4}q_n^4\Big]
+\sum_{n,n'}\frac{C_{n,n'}}{4}(q_n-q_{n'})^2,
\end{equation}
with some coupling coefficients $C_{n,n'}=C_{n',n}\ll \omega_0$. In this case we can introduce
the normal complex variables $a_n=(\tilde q_n+i\tilde p_n)/\sqrt{2}$ through the 
weakly nonlinear canonical transform
\begin{eqnarray}
q_n&=&\tilde q_n\Big(1-\frac{5\varkappa}{32\omega_0}\tilde q_n^2
-\frac{9\varkappa}{32\omega_0}\tilde p_n^2\Big)+\cdots,\\
p_n&=&\tilde p_n\Big(1+\frac{15\varkappa}{32\omega_0}\tilde q_n^2
+\frac{3\varkappa}{32\omega_0}\tilde p_n^2\Big)+\cdots,
\end{eqnarray}
where the dots mean higher-order terms.
As the result, the new Hamiltonian is approximately
\begin{eqnarray}
\tilde{\mathsf H}&\approx&\sum_n\Big(\omega_0|a_n|^2+\frac{g}{2}|a_n|^4\Big)\nonumber\\
&+&\sum_{n,n'}\frac{C_{n,n'}}{8}(a_n+a^*_n-a_{n'}-a^*_{n'})^2,
\end{eqnarray}
with $g=3\varkappa/4$. It is easy to derive equations of motion from here.
For slow envelopes $A_n(t)=a_n(t)\exp(i\omega_0 t)$, neglecting quickly oscillating terms, 
we obtain
\begin{equation}
i(\dot A_n+\gamma A_n)\approx g|A_n|^2A_n +\sum_{n'}\frac{C_{n,n'}}{2}(A_n-A_{n'}).
\label{A_n_eq}
\end{equation}
Now we assume that multi-index $n$ corresponds to nodes of a lattice in 3D space, 
and  a quasi-continuous long-scale regime exists, when $A_n(t) \longrightarrow A({\bf r},t)$. 
Let the coupling coefficients are of a short range,  so that
\begin{equation} 
\sum_{n,n'}C_{n,n'}|A_n-A_{n'}|^2 \longrightarrow \int F({\bf r})|\nabla A|^2d{\bf r}.
\end{equation}
Then in Eq.(\ref{A_n_eq})
$\sum_{n'}C_{n,n'}(A_n-A_{n'}) \longrightarrow -\nabla\cdot (F({\bf r})\nabla A)$,
and we arrive at Eq.(\ref{A_eq}).

Unfortunately, an experimental realization of the above model, with arbitrary dispersion 
$F({\bf r})$ ``on demand'', does not exist yet. But it seems possible in the future 
as an artificially created compound material consisting of a 3D array of nearly identical 
nonlinear oscillators coupled by tuned space-dependent weak links. Ideally, there could 
be full controllability on each individual oscillator and each coupling coefficient. 
Such a durable product could operate at not very low temperatures and be used repeatedly, 
in contrast to a BEC. Physically, the oscillators could be optical, electronic, electromagnetic 
resonators, or something else. That is of course a very important point, but here we do not 
concentrate on the experimental side; only some mathematical aspects of the problem are touched.
Since mathematics is quite interesting in this case, the absence of experimental results 
in this work may hopefully be excused, to some extent.

Thus, although Eq.(\ref{A_eq}) is more suitable for analytic studies, it makes sense also 
to consider its spatially discrete variant Eq.(\ref{A_n_eq}). 
It should be mentioned that discrete forms of NLSE with translationally invariant 
coupling coefficients were investigated previously in many works in different context
(see, e.g., Refs. \cite{KMFC2004,CKMF2005,KFCMB2005,XZL2008,CJKL2009}, and citations therein).
Below, after a brief discussion on general properties of the continuous model (\ref{A_eq}) 
in comparison to GPE with a trapping potential, we actually  study numerically 
a discrete equation (\ref{A_n_eq}) for some particular choice of $C_{n,n'}$.
Similarly to the recent results about persistent vortex knots in GPE \cite{R2018-4,TRK2019},
here we observe analogous long-lived structures, but as we will see,
with sufficiently small values of the grid spacing parameter only. 

\section {General remarks about the model}

It is convenient to get rid of the dissipative term in Eq.(\ref{A_eq})
by introducing a new complex field $\psi$ through the following substitution,
\begin{equation}
A({\bf r}, t)=A_0\psi({\bf r}, t)\exp[-\gamma t-i\sigma(t)],
\end{equation}
where real $A_0$ is a typical amplitude at $t=0$.
With an appropriate real function $\sigma(t)$, we obtain a non-autonomous Hamiltonian system
\begin{equation}
i\psi_t=-\frac{1}{2}\nabla\cdot (F({\bf r})\nabla \psi) 
+gA_0^2e^{-2\gamma t}(|\psi|^2-1)\psi,
\label{psi_eq}
\end{equation}
with the Hamiltonian functional
\begin{equation}
H=\frac{1}{2}\int \big[F({\bf r})|\nabla\psi|^2+gA_0^2e^{-2\gamma t}(|\psi|^2-1)^2\big]d^3{\bf r}.
\end{equation}
What is characteristic for this model, the equilibrium vortex-free state is of 
the simple form $\psi_0=1$. However, when a vortex is present, its local core width 
$\tilde\xi$ (the healing length) depends on ${\bf r}$ and $t$ as
\begin{equation}
\tilde \xi({\bf r},t)=[F({\bf r})/gA_0^2]^{1/2}\exp(\gamma t).
\end{equation}
A gradual broadening of the core takes place because in Eq.(\ref{psi_eq}) the effective 
nonlinear coefficient $gA_0^2\exp(-2\gamma t)$ decays with time.

Below we use dimensionless time and length variables determined by a typical value 
$f$ of function $F({\bf r})$ and by a typical spatial scale $l$ where it varies:
$t_{\rm new}=t_{\rm old}\cdot f/l^2$, and ${\bf r}_{\rm new}={\bf r}_{\rm old}/l$.
The new time unit corresponds to a typical vortex turnover period.
As a result, we have two dimensionless parameters in the system,
\begin{equation}
\xi=(f/gA_0^2l^2)^{1/2},\qquad \delta=\gamma l^2/f,
\end{equation}
where $\xi$ is a typical relative width of the vortex core at $t=0$, 
and $\delta$ is the dimensionless damping rate.
For long-lived vortices to be observable, both parameters should be as small as
$\xi\lesssim 1/15$ and $\delta\lesssim 1/40$. Compatibility of these conditions imposes severe 
constraints on the physical damping rate: $(\gamma/gA_0^2)= \xi^2\delta\lesssim 10^{-4}$.
So a Q-factor of the oscillators should be very large, $Q=\omega_0/\gamma\sim 10^5$.

We will assume $F({\bf r})$ to be positive inside some finite domain ${\cal D}$, 
and be zero on the boundary $\partial{\cal D}$. Such a domain constitutes a closed system.

The standard Madelung transform $\psi=\sqrt{\rho}\exp(i\Phi)$ represents model (\ref{psi_eq})
in hydrodynamic form. In particular, the continuity equation is
\begin{equation}
\rho_t +\nabla\cdot(\rho F\nabla\Phi)=0,
\end{equation}
so the velocity field is defined by equality ${\bf v}= F({\bf r})\nabla\Phi$.
A similar continuity equation for appropriately non-dimensionalized GPE, 
\begin{equation}
i\Psi_t=-(1/2)\Delta\Psi+[V({\bf r})-\mu +|\Psi|^2]\Psi,
\end{equation}
is known to be
\begin{equation}
(|\Psi|^2)_t +\nabla\cdot(|\Psi|^2\nabla\Phi)=0.
\end{equation}
We see that vortical states in both systems have many features in common, if the 
background density of GPE is proportional to $F$ of our model,
$|\Psi_0({\bf r})|^2\propto F({\bf r})$, where $\Psi_0({\bf r})$ is the vortex-free ground 
state for GPE. Indeed, far from a vortex core we have 
$\nabla\cdot(F\nabla\Phi)=0$ for the present model, 
and $\nabla\cdot(|\Psi_0|^2\nabla\Phi)=0$ for GPE, 
which conditions are mathematically equivalent. 
Around a vortex, the phase increment is $2\pi$. However, unlike the GPE model,
velocity circulation is not constant in our case. Another important difference 
is that a local healing length in our model is directly proportional to $\sqrt{F}$, 
while in GPE it is inversely proportional to $|\Psi_0|$. 
Despite the differences, we have much similar equations of  motion
for a vortex line in the hydrodynamical limit for both systems. 
Let a central vortex-core line be parametrized by a vector function ${\bf R}(\beta,t)$, 
with an arbitrary longitudinal parameter $\beta$. Then in our model (where at equilibrium 
$\rho=1$) the equation of motion for curve ${\bf R}(\beta,t)$ has a general variational form 
(below symbol $\hat\delta$ means variation; one should distinguish it from the damping rate 
$\delta$ introduced previously) 
\begin{equation}
2\pi[{\bf R}_\beta\times{\bf R}_t]\cdot 1=\hat\delta{\cal H}/\hat\delta{\bf R}.
\label{R_eq}
\end{equation}
It is assumed that in the hydrodynamical limit, the wave field $\psi$
is completely determined by the vortex line configuration, that is
$\psi({\bf r},t)\approx \psi({\bf r},\{{\bf R}(\beta,t)\})$. 
In Refs.\cite{N2004,BN2015},
it is shown how the standard Hamiltonian structure $i\psi_t=\hat\delta H/\hat\delta\psi^*$ 
of Eq.(\ref{psi_eq}) gives Eq.(\ref{R_eq}). The vortex Hamiltonian is
\begin{equation}
{\cal H}\{{\bf R}\}=H\{\psi({\bf r},\{{\bf R}\}), \mbox{c.c.}\}.
\end{equation}
The same approach as in Refs.\cite{N2004,BN2015}, but with taking into account spatial 
nonuniformity of the ground density, gives for vortex line in GPE model the equation
\begin{equation}
2\pi[{\bf R}_\beta\times{\bf R}_t]|\Psi_0({\bf R})|^2=\hat\delta{\cal H}_{GPE}/\hat\delta{\bf R},
\label{R_eq_GPE}
\end{equation}
(see also Ref.\cite{R2018-1} and references therein for a detailed discussion).
What is important, expressions for the corresponding vortex Hamiltonians ${\cal H}$ and 
${\cal H}_{GPE}$ coincide in the main order on the presumably large parameter 
$\Lambda=\ln(1/\xi)\gg 1$.  Indeed, they are mainly determined by the terms  
$(1/2)\int F|\nabla\Phi|^2 d{\bf r}$ and $(1/2)\int |\Psi_0({\bf R})|^2|\nabla\Phi|^2 d{\bf r}$ 
respectively.
In particular, the local induction approximation (LIA) 
for a single distorted vortex ring is given by the formula
\begin{equation}
{\cal H} \approx \pi[\Lambda -\delta t]\oint  F({\bf R}) |{\bf R}_\beta|d\beta,
\qquad [\Lambda -\delta t]\gg 1.
\label{H_LIA}
\end{equation}
Here the factor $[\Lambda -\delta t]$ is the logarithm of the inverse typical  vortex core
width. The width depends on time since the dissipation is present.
For GPE, the corresponding expression is 
\begin{equation}
{\cal H}_{GPE} \approx \pi\Lambda \oint  |\Psi_0({\bf R})|^2 |{\bf R}_\beta|d\beta,
\qquad \Lambda \gg 1.
\end{equation}
Combining Eq.(\ref{R_eq}) and Eq.(\ref{H_LIA}), we obtain an explicit approximate 
equation of motion for a vortex ring (with some particular longitudinal parameterization),
\begin{equation}
{\bf R}_t=\frac{[\Lambda -\delta t]}{2}\big\{F({\bf R})\kappa {\bf b}
+[\nabla F({\bf R})\times {\bf t}]\big\},
\label{R_t_LIA}
\end{equation}
where ${\bf t}$ is the local unit tangent vector along the curve, ${\bf b}$ is the unit 
binormal vector, and $\kappa$ is the line curvature. This sort of LIA equation is a new result.
Unfortunately, Eq.(\ref{R_t_LIA}) is insufficient to describe vortex knots and links
because non-local interactions are known to be crucially important for them. Besides that,
actual values of $\Lambda$ are not very large in many interesting cases (so, in our numerical
simulations described below, $\Lambda\approx 3$).
Nevertheless, some more words about LIA will be in place here.
Since we are mainly interested in axially symmetric $F=F(z,r)$, with $r=\sqrt{x^2+y^2}$,
it is convenient to parametrize a ring shape by azimuthal angle $\varphi$ 
in the cylindrical coordinates as $z(\varphi,t)$ and $r(\varphi,t)$. 
In that case we have the following system of two scalar equations,
\begin{eqnarray}
rz_t&=&\frac{[\Lambda -\delta t]}{2}[F_r S  -\partial_\varphi(F r_\varphi/S)+rF/S],
\\
-rr_t&=&\frac{[\Lambda -\delta t]}{2}[F_z S -\partial_\varphi(F z_\varphi/S)],
\end{eqnarray}
where $S\equiv\sqrt{r^2+r_\varphi^2+z_\varphi^2}$. For example, if
\begin{equation}
F=\mbox{max}\{(3/2-(r^2+\lambda^2z^2)/2),0\},
\label{Fzr}
\end{equation}
with an anisotropy parameter $\lambda$, then there exists a stationary perfect-ring 
solution $r=1$, $z=0$. For small ring distortions, the linearized equations
of motion for $m$-th azimuthal mode take form
\begin{eqnarray}
 \dot z_m&=&\frac{[\Lambda -\delta t]}{2}(m^2-3)r_m,
\\
-\dot r_m&=&\frac{[\Lambda -\delta t]}{2}(m^2-\lambda^2)z_m.
\end{eqnarray}
These equations are easily solved as
\begin{equation}
\sqrt{|m^2-3|}r_m-i\sqrt{|m^2-\lambda^2|}z_m=C_m e^{-i\Omega_m\tau(t)},
\end{equation}
where $C_m$ are arbitrary complex constants, and
\begin{equation}
\Omega_m=\mbox{sgn}(m^2-3)\sqrt{(m^2-3)(m^2-\lambda^2)},
\end{equation}
\begin{equation}
\tau(t)=[\Lambda t-\delta t^2/2]/2.
\end{equation}
This solution has many features in common with the analogous solution for GPE 
\cite{ring_instability,R2017-3,R2018-1}.

Returning again to more general Eqs. (\ref{R_eq}) and (\ref{R_eq_GPE}), 
it is important to note that recently found long-lived torus vortex knots 
and links in trapped BECs exist in a region where the condensate density 
$|\Psi_0({\bf R})|^2$ varies relatively weakly.
That is why one can expect qualitatively similar behavior of vortex knots in the present 
model if we take for $F$ nearly the same axially symmetric profile Eq.(\ref{Fzr})
as it was for the corresponding density of anisotropic harmonically trapped 
BEC \cite{R2018-4,TRK2019}.
To check this hypothesis, we performed numerical experiments using
a discrete variant of Eq.(\ref{psi_eq}).

\begin{figure}
\begin{center}
\epsfig{file=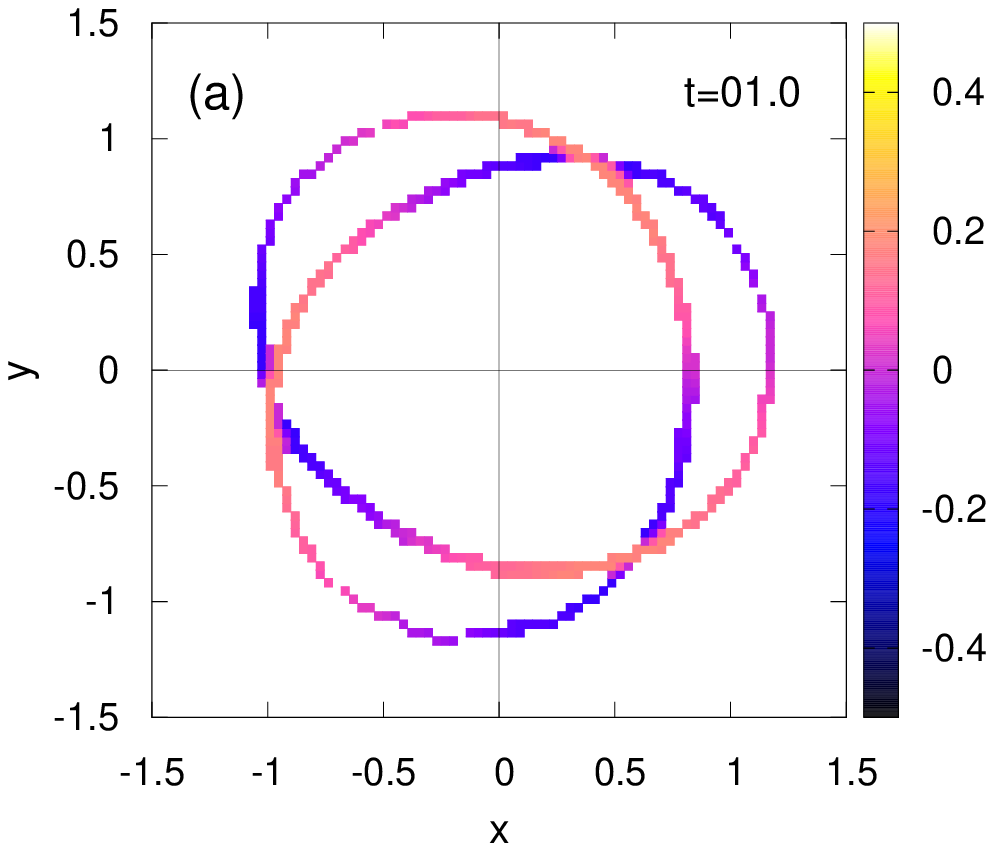, width=59mm}\\
\epsfig{file=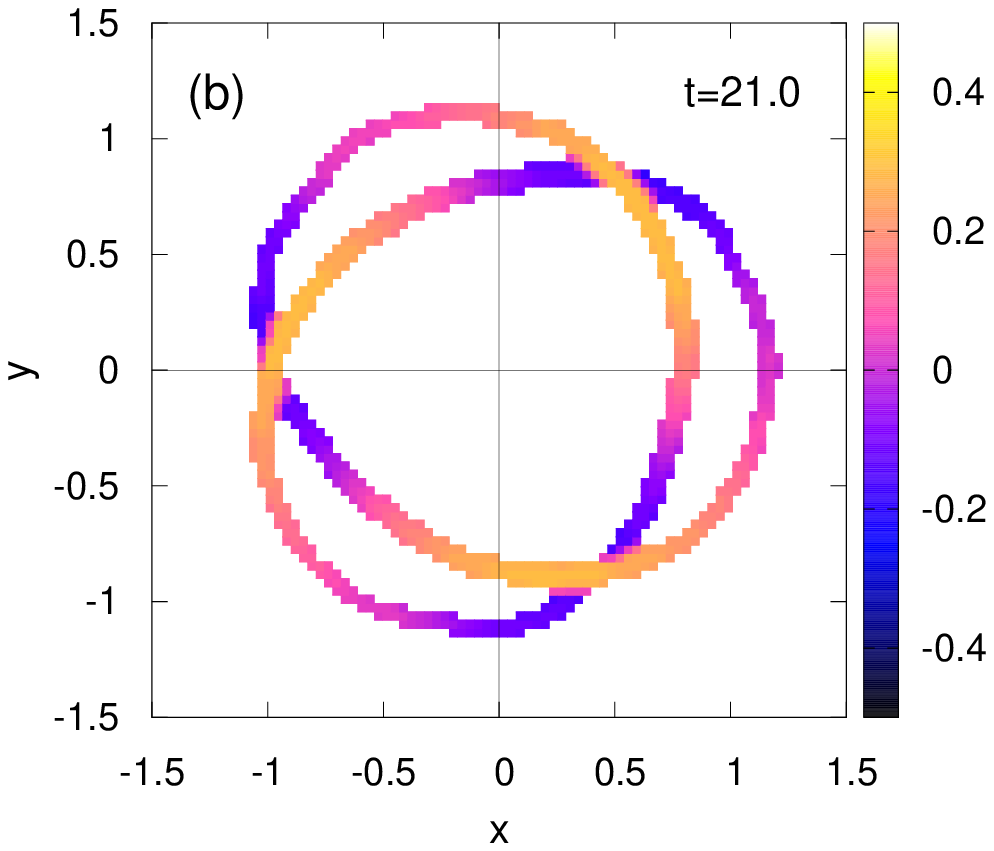, width=59mm}\\
\epsfig{file=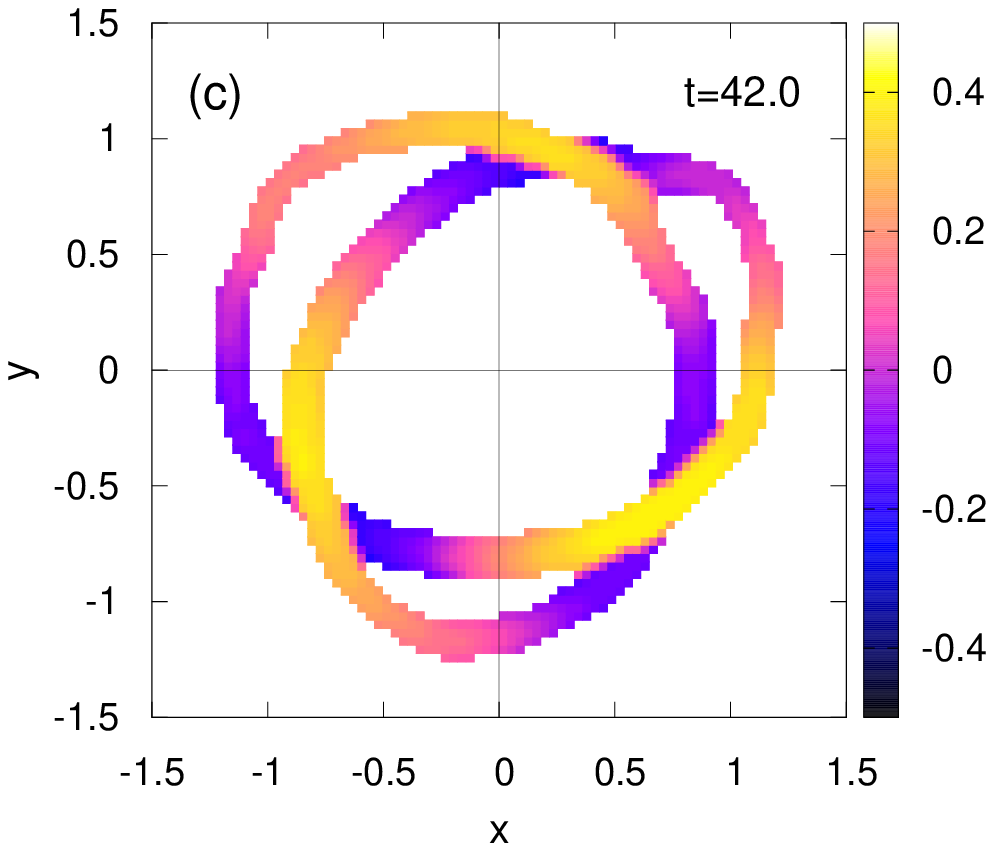, width=59mm}\\
\epsfig{file=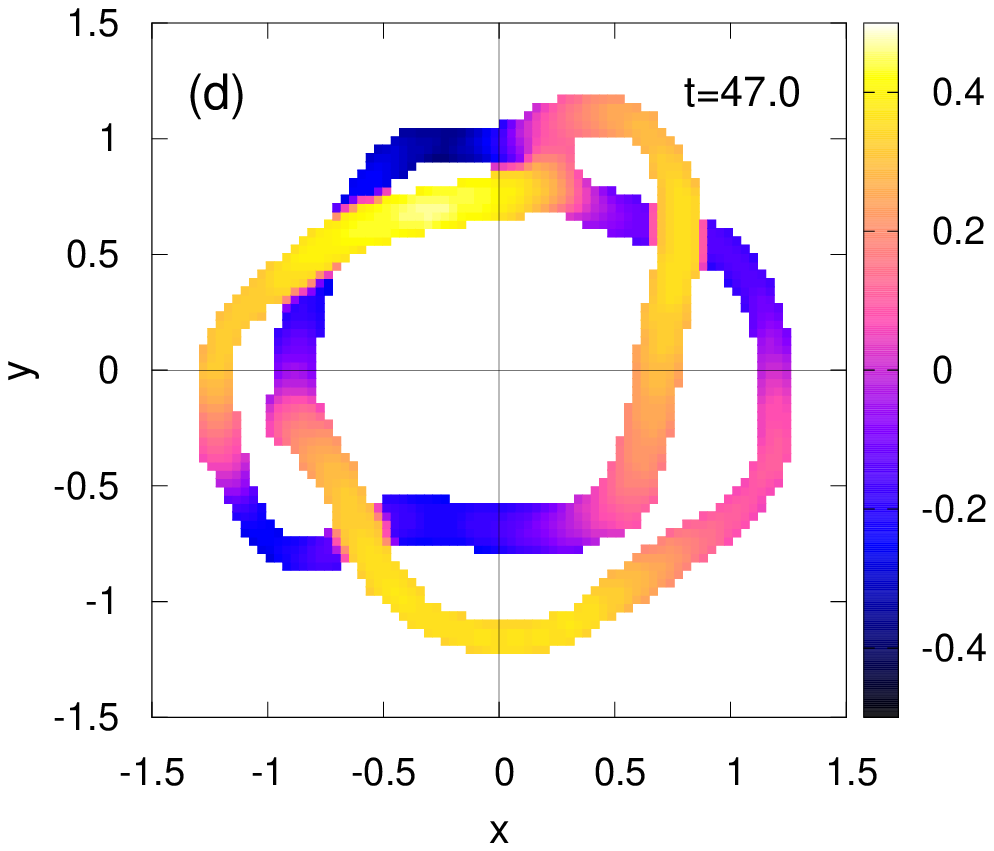, width=59mm}
\end{center}
\caption{An example of evolution of a vortex trefoil knot on a cubic lattice. 
Parameters of this simulation are:
$\lambda^2=2.5$, $1/\xi=24.0$, $h=0.036$, $\delta=0.02$,  $B_0=0.17$.
Shown is an effective  ``surface'' of the vortex core where the 
density takes values near $\rho =0.5$. The color scale indicates $z$-coordinate.}
\label{vortex1} 
\end{figure}

\begin{figure}
\begin{center}
\epsfig{file=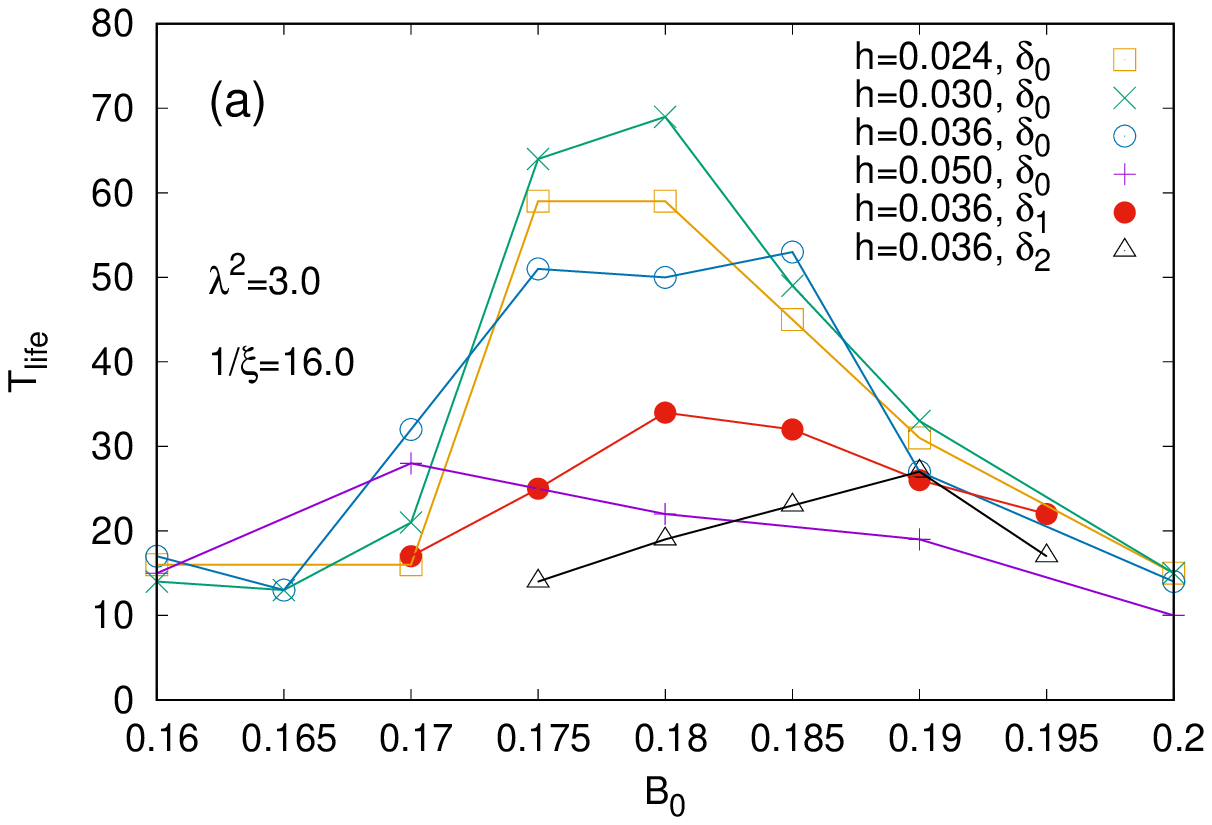, width=82mm}\\
\epsfig{file=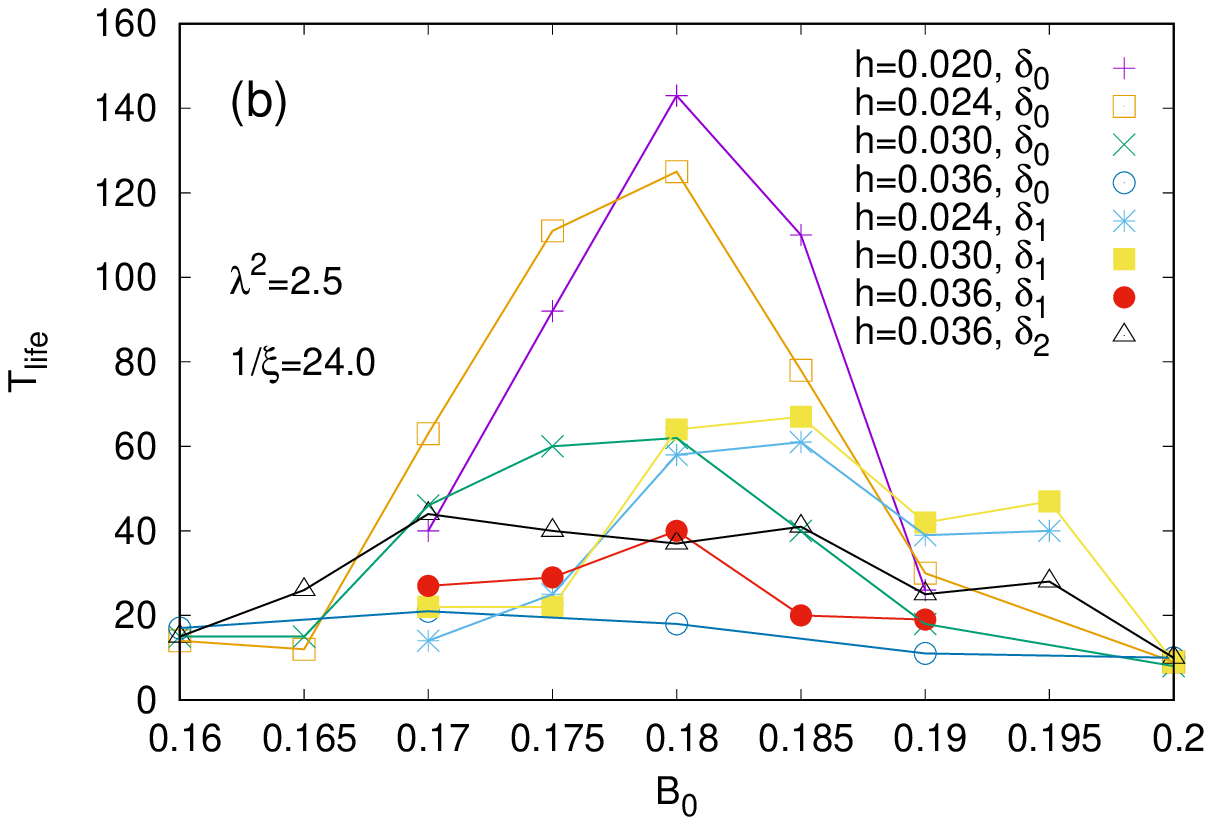, width=82mm}
\end{center}
\caption{Lifetimes of trefoil knot, determined by the moment of first reconnection, 
for different parameters of system (\ref{discrete}).}
\label{lifetimes} 
\end{figure}

\section{Discrete system}

A natural discrete approximation for Eq.(\ref{psi_eq}) corresponds to a simple 
finite-difference scheme with a grid spacing $h\ll 1$. Let nodes of a cubic lattice 
be numbered by an integer vector ${\bf n}=(n_1,n_2,n_3)$, so that ${\bf r}_{{\bf n}}=h{\bf n}$. 
And let the unit basis vectors be ${\bf i}$, ${\bf j}$, and ${\bf k}$.
Then we have a 3D array of undamped coupled oscillators, with space-dependent 
coupling coefficients between the nearest neighbors, 
and with a nonlinear frequency shift exponentially varying in time
(compare to Refs.\cite{KMFC2004,CKMF2005,KFCMB2005,XZL2008,CJKL2009}):
\begin{eqnarray}
i\dot\psi_{\bf n}=\frac{1}{2h^2}\big(
 [\psi_{\bf n}-\psi_{{\bf n}+{\bf i}}]F_{{\bf n}+{\bf i}/2}
+[\psi_{\bf n}-\psi_{{\bf n}-{\bf i}}]F_{{\bf n}-{\bf i}/2}
\big)&&\nonumber\\
+\frac{1}{2h^2}\big(
 [\psi_{\bf n}-\psi_{{\bf n}+{\bf j}}]F_{{\bf n}+{\bf j}/2}
+[\psi_{\bf n}-\psi_{{\bf n}-{\bf j}}]F_{{\bf n}-{\bf j}/2}
\big)&&\nonumber\\
+\frac{1}{2h^2}\big(
 [\psi_{\bf n}-\psi_{{\bf n}+{\bf k}}]F_{{\bf n}+{\bf k}/2}
+[\psi_{\bf n}-\psi_{{\bf n}-{\bf k}}]F_{{\bf n}-{\bf k}/2}
\big)&&\nonumber\\
+[\exp(-2\delta t)/\xi^2](|\psi_{\bf n}|^2-1)\psi_{\bf n},\qquad \qquad&&
\label{discrete}
\end{eqnarray}
where $F_{{\bf n}+{\bf i}/2}=F(h[{\bf n}+{\bf i}/2])$, and so on. Note that interaction 
coefficients are equal to zero if the corresponding midpoints are outside the ellipsoid 
$r^2+\lambda^2z^2=3$. 
So we have a compact structure with a finite number of interacting degrees of freedom 
depending on $h$ and $\lambda$. 

\begin{figure}
\begin{center}
\epsfig{file=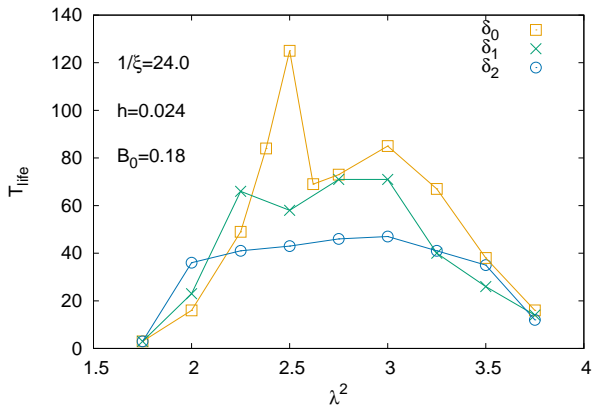, width=82mm}
\end{center}
\caption{Lifetimes of trefoil knot over the anisotropy parameter, for different damping rates.}
\label{TvsA} 
\end{figure}

The above dynamical system has been numerically simulated using
a 4th-order Runge-Kutta time stepping. In our numerical experiments, we took parameter values 
mainly from the following sets: $h=h_{1,2,3,4,5}=\{0.02,0.024,0.03,0.036,0.05\}$; 
$\delta=\delta_{0,1,2}=\{0.00,0.01,0.02\}$; $1/\xi=\{16.0, 24.0\}$; $\lambda^2=\{2.5, 3.0\}$.
The initial vortex shape was a torus trefoil knot. 
When parametrized in the cylindrical coordinates, it is
\begin{equation}
r(\varphi)-i z(\varphi)=1+B_0\exp(3i\varphi/2).
\end{equation}
We empirically found that optimal values for $B_0$ lie within interval $0.16<B_0<0.20$.

To reduce undesirable potential perturbations in the initial state, a special procedure
was used, similar to that described in Refs.~\cite{R2018-4,TRK2019}. Basically, it was
an imaginary-time propagation in a dissipative regime, with a temporarily added pinning 
potential along the prescribed vortex core.

Some numerical results are presented in the figures. In particular, Fig.1 gives an example 
where, for $\xi\sim h$, the knot preserves its topology over more than 40 time units, 
and then reconnection occurs as the result of growing shape perturbations, 
much like in the GPE simulations \cite{R2018-4,TRK2019}. 
The core width is seen to gradually increase with time, in accordance with the theory.

Fig.2 exhibits collected information about trefoil lifetimes for different parameters.
First of all, it should be noted that grid spacing $h$ has a strong influence on vortex dynamics.
In particular, for $\delta=0$ there are crossover regions in $h$ from a strictly discrete regime 
to a quasi-continuous regime. In quasi-continuous regime and with small $\xi$, the lifetimes
are very large, more than a hundred of vortex turnover periods. On the other hand,
the discrete regime, with a coarse grid $h\sim \xi$, is not favorable for vortex.
With a finite damping rate $\delta$, situation is more interesting, since the crossover can occur
dynamically during the evolution. Fig.1 illustrates such a case.
At the beginning we have the discrete regime, but after a time, 
the core width becomes more ``fat'', $\xi\exp(\delta t)\gtrsim 2 h$, so the vortex enters 
quasi-continuous regime, however with not so thin core as is required for a very long lifetime. 
An overall result is determined by interplay between the two opposite factors, 
the quasi-continuous regime (good for vortex), and fat core (bad for vortex). 
Therefore in some cases a larger value of $\delta$ results in a longer lifetime,
and this somewhat paradoxical result takes place entirely due to the discreteness. 

Finally, Fig.\ref{TvsA} shows how the lifetime depends on the anisotropy parameter.
It is seen that there exists an optimal interval for $\lambda^2$ where the knot is long-lived. 

\section{Conclusion}

In this work, an alternative with respect to the Gross-Pitaevskii equation 
theoretical model has been suggested (namely a NLSE with a space-varying dispersion) 
where enhanced lifetimes of quantized vortex knots are possible. 
There is a hope that possible physical implementations of the model will potentially 
have practical advantages compared with Bose-Einstein condensates, 
as vortex dynamics is investigated. In numerical simulations we observed that a discrete 
NLSE with space-dependent coupling coefficients can demonstrate solutions having 
very interesting properties. In particular, our numerical experiments have confirmed 
the idea that vortex knots on a lattice can exist for relatively long times.

In perspective, generalization of the present model to the case when $F({\bf r})$ 
is a matrix function seems even more promising, since vortices can have unusual 
local and global anisotropic properties in such systems.

\end{document}